\documentclass{article} 
\usepackage{nips12submit_e,times}

\usepackage{graphics} 
\usepackage{amsmath} 
\usepackage{amssymb}  
\usepackage{cite}
\usepackage{ gensymb }
\usepackage{bm}
\usepackage{subfigure}
\usepackage[pdftex]{graphicx}
\graphicspath{{figures/}}
\DeclareGraphicsExtensions{.pdf,.jpeg,.png}

\newtheorem{theorem}{Theorem}[section]

\newtheorem{definition}[theorem]{Definition}

\newtheorem{problem}[theorem]{Problem}

\title{An Information Value Function for Nonparametric Gaussian Processes}

\author{
H.Wei \\
Department of Mechanical Engineering and Materials Science\\
Duke University\\
Durham, NC 27708\\
\texttt{hongchuan.wei@duke.edu} \\
\And
W.Lu \\
Department of Mechanical Engineering and Materials Science\\
Duke University\\
Durham, NC 27708\\
\texttt{wenjie.lu@duke.edu} \\
\AND
S. Ferrari\\
Department of Mechanical Engineering and Materials Science\\
Duke University\\
Durham, NC 27708\\
\texttt{sferrari@duke.edu}
}

\nipsfinalcopy 

\begin{document}

\maketitle

\begin{abstract}
This paper presents a novel information value function that can be used in online sensor planning to monitor a spatial phenomenon in which the spatial phenomenon is modeled by nonparametric Gaussian processes. The information value function is derived from the Kullback-Leibler (KL) divergence and represents the information value brought by sensor decision. The sensor decision at every time step is to select the sensing location that maximizes the information value function associated with the measurement taken. Gaussian processes (GPs) are employed to obtain the posterior distribution of the spatial phenomenon given a number of sensor measurements, because GPs have sufficient flexibilities to adopt the complexity from data. Furthermore, a greedy algorithm is designed based on the information value function. By comparing the greedy algorithm with the random algorithm, it is shown that the error decreases faster defined as the difference between the estimated posterior distribution and the true distribution of the spatial phenomenon via the greedy algorithm.
\end{abstract}

\section{Introduction}
\label{sec:Introduction}
The problem of monitoring spatial phenomena \cite{GuestrinNearOptimalSensorPlacement05} with little or no prior information is relevant to a variety of applications,including monitoring the atmospheric temperature for the conterminous United States\cite{DalyTemperature02}. In this context, sensor planning can be viewed as a decision making problem in which the sensor is an information-gathering agent that decides its measurement sequence in order to optimize the sensing performance over time. Sensor planning algorithms require a closed form representation of the sensing performance as a function of the measurement sequence and the spatial phenomenon distribution. In this paper, we show that under proper assumptions, monitoring a spatial phenomenon can be reduced to the problem of estimating a probability density function (PDF) representation of the spatial phenomenon given partial or imperfect sensor measurements \cite{SchmaedekeInformationBasedSensorManagement93}. Then, the sensing performance can be expressed by certain metrics of the PDF representing the spatial phenomenon, such as the information entropy.

Information value functions have been used to quantify the amount of information associated with random variables such as the values of the spatial phenomenon at a set of targets, and to control or manage sensor measurements to minimize the uncertainty of the spatial phenomenon\cite{ZhangInformationRoadmapSensorPathPlanning09,CaiDemining09}. Computing information value functions for one or more random variables requires knowledge of the random variables' joint probability mass (or density) functions. Therefore, in order to derive an information value function associated with a measurement sequence, the corresponding posterior distribution (also known as the posterior belief) of the spatial phenomenon given the measurement sequence is derived and utilized. A general approach was recently presented by the authors for estimating the \emph{expected} information value of future sensor measurements in target classification problems\cite{ZhangComparisonInformationFunctions12}. This paper extends the approach in \cite{ZhangComparisonInformationFunctions12} to monitoring spatial phenomena modeled by GPs. In \cite{ZhangComparisonInformationFunctions12}, the target classification problem has a discrete classification space with a limited size, while the spatial phenomenon in this paper is defined in a continuous domain. Therefore, if the same technique in \cite{ZhangComparisonInformationFunctions12} was applied to monitoring the spatial phenomenon, discretization of the continuous domain would result in algorithms with a high computational complexity. Gaussian processes are adopted in this paper to model the spatial phenomenon and to resolve the problem raised by the continuous domain, since GPs can simplify the estimation of the spatial phenomenon to closed form functions given noisy measurements. Then, the KL-divergence can be utilized to derive the information value function, which generates the optimal sensor action given previous measurements. The advantage of the proposed approach includes that the information theoretic function can be evaluated at any set of points within the function domain without discretization.

The paper is organized as follows. The problem is formulated in Section \ref{sec:ProblemFormulation}.
Section \ref{sec:background} gives the background knowledge of Gaussian processes. The methodology is
given and analyzed in Section \ref{sec:methodology}. The simulation results are presented in Section
\ref{sec:Results}. At last, the conclusions are drawn in Section \ref{sec:Conclusion}.

\section{Problem Formulation}
\label{sec:ProblemFormulation}
The sensor planning problem considered in this paper involves managing a measurement sequence of a sensor for the purpose of monitoring a time-invariant spatial phenomenon $f$ in a finite-size region of interest (RoI), $ \mathcal{A} \subset \mathbb{R}^2$. We assume that the RoI is a connected subspace of two-dimensional Euclidean space, that is, $\mathcal{A}$ cannot be divided into two disjoint nonempty closed sets. A set of targets is distributed within the RoI, and is organized in the set,
\begin{equation}
	\mathcal{V} = \{ \bm{\alpha}_i | i=1,\cdots,N_v \},\quad \bm{\alpha}_i \in \mathcal{A},
\end{equation}
where the targets in $\mathcal{V}$ represent points in $\mathcal{A}$ of higher interest than points everywhere in $\mathcal{A}\backslash\mathcal{V}$. The '$\backslash$' denotes the complementary set. By comparing the posterior distribution of the targets in $\mathcal{V}$ given the measurement sequence and the true values of the spatial phenomenon $f$ at $\mathcal{V}$, we can evaluate the performance of the sensor planning strategy. Notice that the set $\mathcal{V}$ is not necessary stationary over time, i.e., at each time step, users can specify different targets due to the change of external environment or the interest. Additionally, the size of $\mathcal{V}$ can vary over time.

\begin{figure}[h]
    \centering
    \includegraphics[width=5in]{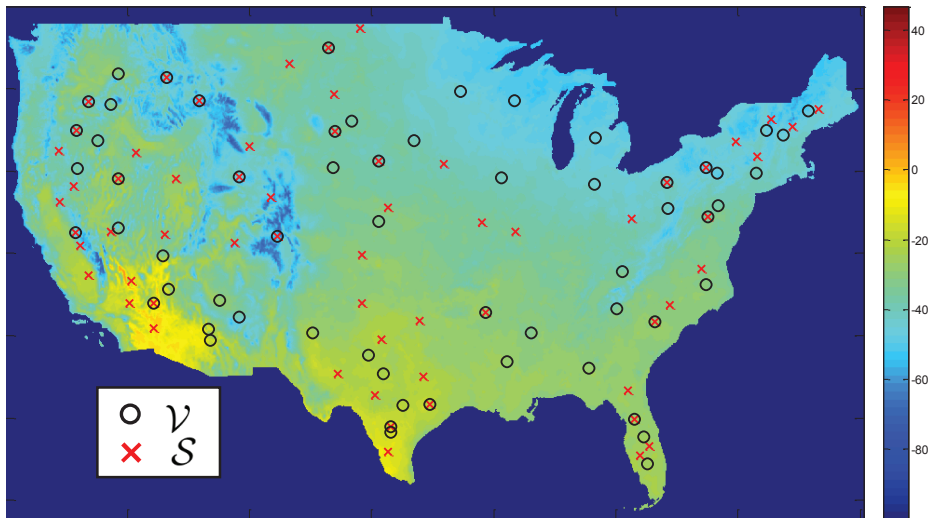}
    \caption{Example of RoI, $\mathcal{A}$, showing the conterminous Unites States, over which the temperature data analyzed in this paper is defined. A set of targets, $\mathcal{V}$ of size $N_v=61$ is denoted by black circles. A set of accessible sensing positions, $\mathcal{S}$, of size $N_s=60$ is denoted by red crosses. The intersection of the set of targets and the set of accessible sensing positions, $\mathcal{I}$, is denoted by a cross in the circle.}
    \label{fig:RoI}
\end{figure}

The spatial phenomenon $f$ is modeled as a Gaussian process, and the GP is updated given the measurements obtained by the sensor online. At each time step, the sensor takes a measurement of $f$ at $\mathbf{x}$, where $\mathbf{x}$ is selected from $N_s$ accessible sensing locations scattered in the RoI, that is, $\mathcal{S}=\{\mathbf{s}_i|i=1,\cdots,N_s \} \subset \mathcal{A}$. The set of accessible sensing positions, $\mathcal{S}$, are known \textit{a priori}, however, no restrictions on its distribution need to be made. The size of the set of accessible sensing positions, $N_s$, is limited due to the low control accuracy of the sensor actuator, for example, the sensor can only receive signals from certain directions. Fig. \ref{fig:RoI} illustrates an example of a RoI with an irregular shape, populated with a set of targets and a set of accessible sensing positions. Notice that the locations in the set $\mathcal{V}$ do not have to be selected from the set $\mathcal{S}$.

At the $k$th time step, the sensor takes one measurement at one location $\mathbf{y}_k \in \mathcal{S}$ and returns a noisy measurement, $z_k$, which is modeled as
\begin{equation}
\label{eq:SensorModel}
	z_k = f_0(\mathbf{y}_k)+\varepsilon,
\end{equation}
where $\mathbf{y}_k$ is the $k$th sensing location chosen by the algorithm introduced in Section \ref{sec:methodology}, $f_0(\mathbf{y}_k)$ is the true value of the spatial phenomenon at $\mathbf{y}_k$, $\varepsilon$ is an additive, zero-mean Gaussian noise, which is fully specified by its standard deviation $\sigma$. After the sensor takes its measurement at the $k$th time step, the posterior distribution $\hat{f}(\mathbf{x}|\mathbf{Y}_k,\mathbf{Z}_k)$ of the spatial phenomena $f$ is updated, where $\mathbf{Y}_k=\{\mathbf{y}_1,\dots,\mathbf{y}_k\}$ denotes the set of chosen sensing locations, and $\mathbf{Z}_k=\{{z}_1,\dots,{z}_k\}$ denotes the set of taken measurements. Then the sensor planning problem can be stated as Problem \ref{prob:SensorPlanning}.

\begin{problem}\label{prob:SensorPlanning}
Find $\mathbf{y}_1,~\mathbf{y}_2,~\cdots$ to minimize
\begin{equation}
\label{eq:SensorPlanningProblem}
\sum\limits_{\mathbf{x}_i \in \mathcal{V}}(f_0(\mathbf{x}_i)-\hat{f}(\mathbf{x}_i|\mathbf{Y}_k,\mathbf{Z}_k))^2.
\end{equation}
subject to the sensor model (\ref{eq:SensorModel})
\end{problem}
In the Section \ref{sec:methodology}, the information value function based on the KL-divergence is developed to choose the measurement sequence $\mathbf{y}_1,~\mathbf{y}_2,~\cdots$.

\section{Background on Gaussian Processes}
\label{sec:background}
A Gaussian process is a nonparametric Bayesian model, and is a distribution defined over functions, $f(\mathbf{x})$, where $f(\mathbf{x})$ is a function mapping the input space $\Omega_{\mathbf{x}}$ to $\mathbb{R}$,
\begin{equation}\label{eq:GPDefinition}
	f(\mathbf{x}):\Omega_{\mathbf{x}} \rightarrow \mathbb{R}, \quad \mathbf{x} \in \Omega_{\mathbf{x}}.
\end{equation}
\begin{definition}
A Gaussian process is a collection of random variables, any finite number of which have a joint Gaussian distribution.
\end{definition}
For a rigorous definition and a comprehensive review of Gaussian processes, the reader is referred to \cite{RasmussenWilliamsGPforMachineLearning06}.

Gaussian processes are widely used to model spatial phenomena due to its ability to deal with large data set and to recruit more parameters as the size of the data grows \cite{CampbellNPBM12}. Furthermore, GPs return the posterior estimation of spatial phenomena in closed form, which reduces the computational complexity in a number of applications. A GP is completely specified by its mean function $m(\mathbf{x})$ and covariance function $k(\mathbf{x}_1,\mathbf{x}_2)$, such that
\begin{equation}
\label{eq:GPMeanCovariance}
	\begin{aligned}
	m(\mathbf{x}) & = E[f(\mathbf{x})], \\
	k(\mathbf{x}_1,\mathbf{x}_2) &=E[(f(\mathbf{x}_1)-m(\mathbf{x}_1))(f(\mathbf{x}_2)-m(\mathbf{x}_2))],\\
	\end{aligned}
\end{equation}
where $E[\cdot]$ denotes the expectation. As a consequence, A Gaussian process can be written as
\begin{equation}
	f(\mathbf{x}) \sim \mathcal{GP}(m(\mathbf{x}),k(\mathbf{x}_1,\mathbf{x}_2)).
\end{equation}
If we are interested in the function at certain points $\mathbf{X}=\{\mathbf{x}_1,\dots,\mathbf{x}_r \}$, we can use $\mathbf{f}(\mathbf{X})$ to denote the $r$-dimensional vector of function values,
\begin{equation}
	\mathbf{f}(\mathbf{X}) = [f(\mathbf{x}_1) \quad \cdots \quad f(\mathbf{x}_n)]^T.
\end{equation}
Similarly, the vector of mean functions is
\begin{equation}
	\mathbf{m}(\mathbf{X}) = [m(\mathbf{x}_1) \quad \cdots \quad m(\mathbf{x}_n)]^T,
\end{equation}
and the covariance matrix $\mathbf{K}(\mathbf{X},\mathbf{Y})$ is
\begin{equation}
\mathbf{K}_{ij} = k(\mathbf{x}_i,\mathbf{y}_j), \quad \mathbf{x}_i \in \mathbf{X}, \mathbf{y}_j \in \mathbf{Y}.
\end{equation}
It follows that the distribution of $\mathbf{f}$ is
\begin{equation}
	\mathbf{f}(\mathbf{X}) \sim \mathcal{N} \left( \mathbf{m}(\mathbf{X}), \mathbf{K}[\mathbf{X},\mathbf{X}] \right),
\end{equation}
where $\mathcal{N}(\mu,\Sigma)$ denotes the multivariate Gaussian distribution with mean $\mu$ and covariance $\Sigma$.

The Gaussian process is powerful in predicting the posterior distribution of a spatial phenomenon. For example, given the set of $k$ observations $\mathbf{Z}_k$ at locations $\mathbf{Y}_k$ and the sensor model in (\ref{eq:SensorModel}), the posterior distribution of the function $\hat{f}$ at any set of targets in the RoI can be derived from this relation
\begin{equation}
\label{eq:RelationPriorPosterior}
	\begin{bmatrix}
		\mathbf{Z}_k \\ \hat{\mathbf{f}}(\mathbf{X})
	\end{bmatrix}
	\sim \mathcal{N}
	\left(
		\begin{bmatrix}
			\mathbf{m}(\mathbf{Y}_k) \\ \mathbf{m}(\mathbf{X})
		\end{bmatrix},
		\begin{bmatrix}
			\mathbf{K}(\mathbf{Y}_k,\mathbf{Y}_k)+\sigma^2 I  & \mathbf{K}(\mathbf{Y}_k,\mathbf{X})\\
			\mathbf{K}(\mathbf{X},\mathbf{Y}_k) & \mathbf{K}(\mathbf{X},\mathbf{X})\\
		\end{bmatrix}
	\right).	
\end{equation}
From the conditional distribution of multivariate Gaussian distributions \cite{BertsekasIntroProb06}, the posterior mean and covariance of $\hat{\mathbf{f}}(\mathbf{X})$ are
\begin{equation}\label{eq:PosteriorMeanCovariance}
	\begin{aligned}
	\boldsymbol{\mu}_k = \mathbf{m}(\mathbf{X}) + \mathbf{K}(\mathbf{X},\mathbf{Y}_k)[\mathbf{K}(\mathbf{Y}_k,\mathbf{Y}_k)+\sigma^2 I]^{-1}(\mathbf{Z}_k-\mathbf{m}(\mathbf{Y}_k)), \\
	\Sigma_k = \mathbf{K}(\mathbf{X},\mathbf{X})-\mathbf{K}(\mathbf{X},\mathbf{Y}_k)[\mathbf{K}(\mathbf{Y}_k,\mathbf{Y}_k)+\sigma^2 I]^{-1}\mathbf{K}(\mathbf{Y}_k,\mathbf{X}).\\
	\end{aligned}
\end{equation}
Depending on (\ref{eq:PosteriorMeanCovariance}), an information value function is developed in the following section, and can be evaluated analytically without integrating numerically over the possible measurements.

\section{Information Value Function}
\label{sec:methodology}
In this section, an information value function is developed based on KL divergence to evaluate the expected discrimination gain (EDG) \cite{ZhangComparisonInformationFunctions12,KastellaDiscriminationgain97} by a sensing action, i.e., measuring the value of $f$ at one location from $\mathcal{S}$. The information value of measuring $f$ at $\mathbf{y}_k$ is
\begin{equation}
    \begin{aligned}
	 &\hat{\varphi}_D(\mathcal{V};z_k|\mathbf{y}_{k},\mathbf{Y}_{k-1},\mathbf{Z}_{k-1}) = \\
    &\int \boldsymbol{D}\Bigl(\mathbf{f}\bigl(\mathcal{V}|\mathbf{Y}_{k-1},\mathbf{Z}_{k-1},\mathbf{x}_k,z_k)\bigr)~||~
    \mathbf{f}\bigl(\mathcal{V}|\mathbf{Y}_{k-1},\mathbf{Z}_{k-1}\bigr) \Bigr) f(z_k|\mathbf{Y}_{k-1},\mathbf{Z}_{k-1},\mathbf{x}_k) d z_{k},
    \end{aligned}
\end{equation}
where $\boldsymbol{D}(\cdot||\cdot)$ denotes the Kullback-Leibler divergence \cite{KullbackOnInformationSufficiency51},
\begin{equation}
	\boldsymbol{D}(P||Q)= - \int \log \frac{dQ}{dP} dP.
\end{equation}

Since the prior and posterior distributions of $\mathbf{f}$ are both multivariate Gaussian distributions, the computation of $\hat{\varphi}_D(\mathcal{V};z_k|\mathbf{y}_{k},\mathbf{Y}_{k-1},\mathbf{Z}_{k-1})$ can be simplified, such that,
\begin{equation}\label{eq:SimplifyPhiD}
    \begin{aligned}
    & \hat{\varphi}_D(\mathcal{V};z_k|\mathbf{y}_{k},\mathbf{Y}_{k-1},\mathbf{Z}_{k-1}) = \\
    & \int_{-\infty}^{\infty} \frac{1}{2}( tr(\Sigma_{k-1}^{-1} \Sigma_k) -\ln(\frac{det(\Sigma_k)}{det(\Sigma_{k-1})}) -N_v + (\boldsymbol{\mu}_{k}-\boldsymbol{\mu}_{k-1})^T \Sigma_{k-1}^{-1} (\boldsymbol{\mu}_{k}-\boldsymbol{\mu}_{k-1}) ) \mathcal{N}(\mu_{z_k},\sigma_{z_k}) d z_k,
    \end{aligned}
\end{equation}
where $det(\cdot)$ denotes the determinant of a matrix, $\mu_{z_k}$ is the posterior mean of $z_k$ given $\mathbf{Y}_{k-1}$, $\mathbf{Z}_{k-1}$ and $\mathbf{y}_k$,
\begin{equation}\label{eq:MuZt}
    \mu_{z_k} = m(\mathbf{y}_k) + \mathbf{K}(\mathbf{y}_k, \mathbf{Y}_{k-1})[\mathbf{K}(\mathbf{Y}_{k-1},\mathbf{Y}_{k-1}) + \sigma^2 I]^{-1} (\mathbf{Z}_{k-1} - \mathbf{m}(\mathbf{Y}_{k-1})),
\end{equation}
and $\sigma_{z_k}$ is the posterior standard variance of $z_k$,
\begin{equation}\label{eq:SigmaZt}
    \sigma_{z_k} = k(\mathbf{y}_k,\mathbf{y}_k) - \mathbf{K}(\mathbf{y}_k, \mathbf{Y}_{k-1})[\mathbf{K}(\mathbf{Y}_{k-1},\mathbf{Y}_{k-1}) + \sigma^2 I]^{-1} \mathbf{K}(\mathbf{Y}_{k-1},\mathbf{y}_k).
\end{equation}
It follows that
\begin{equation}\label{eq:SimplifyPhiDFinal}
    \begin{aligned}
    \hat{\varphi}_D(\mathcal{V};z_k|\mathbf{y}_{k},\mathbf{Y}_{k-1},\mathbf{Z}_{k-1})
    & = \frac{1}{4} d \sigma_{z_k}^3 \sqrt{\pi} + \frac{1}{2} \sigma_{z_k}\sqrt{\pi} \Bigl(tr(\Sigma_{k-1}^{-1} \Sigma_k) -\ln(\frac{det(\Sigma_k)}{det(\Sigma_{k-1})}) -N_v \\
    & + \mathbf{V}_1^T \mathbf{M}_1^T \Sigma_{k-1}^{-1} (\mathbf{M}_1 \mathbf{V}_1 -2 \mathbf{M}_2 \mathbf{V}_2)
      + \mathbf{V}_2^T \mathbf{M}_2^T \Sigma_{k-1}^{-1} \mathbf{M}_2 \mathbf{V}_2 \Bigr), \\
    \end{aligned}
\end{equation}
where
\begin{equation}
    \begin{aligned}
    &\mathbf{M}_1 = \mathbf{K}(\mathcal{V},\mathbf{Y}_{k-1})(\mathbf{K}(\mathbf{Y}_{k-1},\mathbf{Y}_{k-1})+\sigma^2I)^{-1} \\
    &\mathbf{M}_2 = \mathbf{K}(\mathcal{V},\mathbf{Y}_{k})(\mathbf{K}(\mathbf{Y}_k,\mathbf{Y}_k)+\sigma^2I)^{-1} \\
    &\mathbf{V}_1 = \mathbf{Z}_{k-1}-\mathbf{m}(\mathbf{Y}_{k-1}) \\
    &\mathbf{V}_2 = [ (\mathbf{Z}_{k-1}-\mathbf{m}(\mathbf{Y}_{k-1}))^T \quad \mu_{z_k}-m(\mathbf{y}_k)]^T , \\
    \end{aligned}
\end{equation}
$tr(\cdot)$ denotes the trace of a matrix, and the constant $d$ is the last component of matrix $ \mathbf{M}_2^T \Sigma_{k-1}^{-1} \mathbf{M}_2$.

After the information value is calculated for each sensing location $\mathbf{x}\in \mathcal{S}$, the spatial phenomenon at the location with the highest information value is measured.  Thereafter, the estimation of the spatial phenomenon $\hat{f}(\mathbf{x}|\mathbf{Y}_k,\mathbf{Z}_k)$ is calculated given the new measurement \{$z_k$, $\mathbf{y}_k$\}, and previous estimation $\hat{f}(\mathbf{x}|\mathbf{Y}_{k-1},\mathbf{Z}_{k-1})$. Since the measurement sequence is decided one at a time, the algorithm is a greedy algorithm and is summarized in Fig. \ref{fig:TheGreedyAlgorithm}.
\begin{figure}[thpb]
      \centering
      \framebox{\parbox{6in}{
      \textbf{Input:} functions: $\mathbf{m}$, $\mathbf{K}(\cdot,\cdot)$; sets: $\mathcal{S}$, $\mathcal{V}$; scalars: maximum number of observations, $N_f$\\
      \textbf{Output:} sensing location sequence $\mathbf{Y}_{N_f}$ \\
      \textbf{begin}\\
      \hspace*{8 mm}  $\mathbf{Y}_{N_f} \leftarrow \emptyset;$\\
      \hspace*{8 mm}  \textbf{for} $k=1:N_f$\\
      \hspace*{16 mm} $\mathbf{y}_{k} = \underset{\mathbf{y}_k \in \mathcal{S}}{\operatorname{argmax}}  ~ \hat{\varphi}_D(\mathcal{V};z_k|\mathbf{y}_{k},\mathbf{Y}_{k-1},\mathbf{Z}_{k-1})$\\
      \hspace*{16 mm} $\mathbf{Y}[k] = \mathbf{y}_k$\\
      \hspace*{16 mm} $z_k = f_0(\mathbf{y}_k)+\varepsilon$\\
      \hspace*{16 mm} $\mathbf{Z}[k] = z_k$ \\
      \hspace*{8 mm}  \textbf{endfor} \\
      \hspace*{8 mm}  \textbf{return} $\mathbf{Y}$ \\
      \textbf{end} \\
      }}
      \caption{The greedy algorithm based on the information value function}
      \label{fig:TheGreedyAlgorithm}
\end{figure}

\section{Numerical Simulations and Results}
\label{sec:Results}

\begin{figure}[htp]
\centering
\subfigure[Estimation error and variance on $\mathcal{V}$]{
 \includegraphics[scale =0.43] {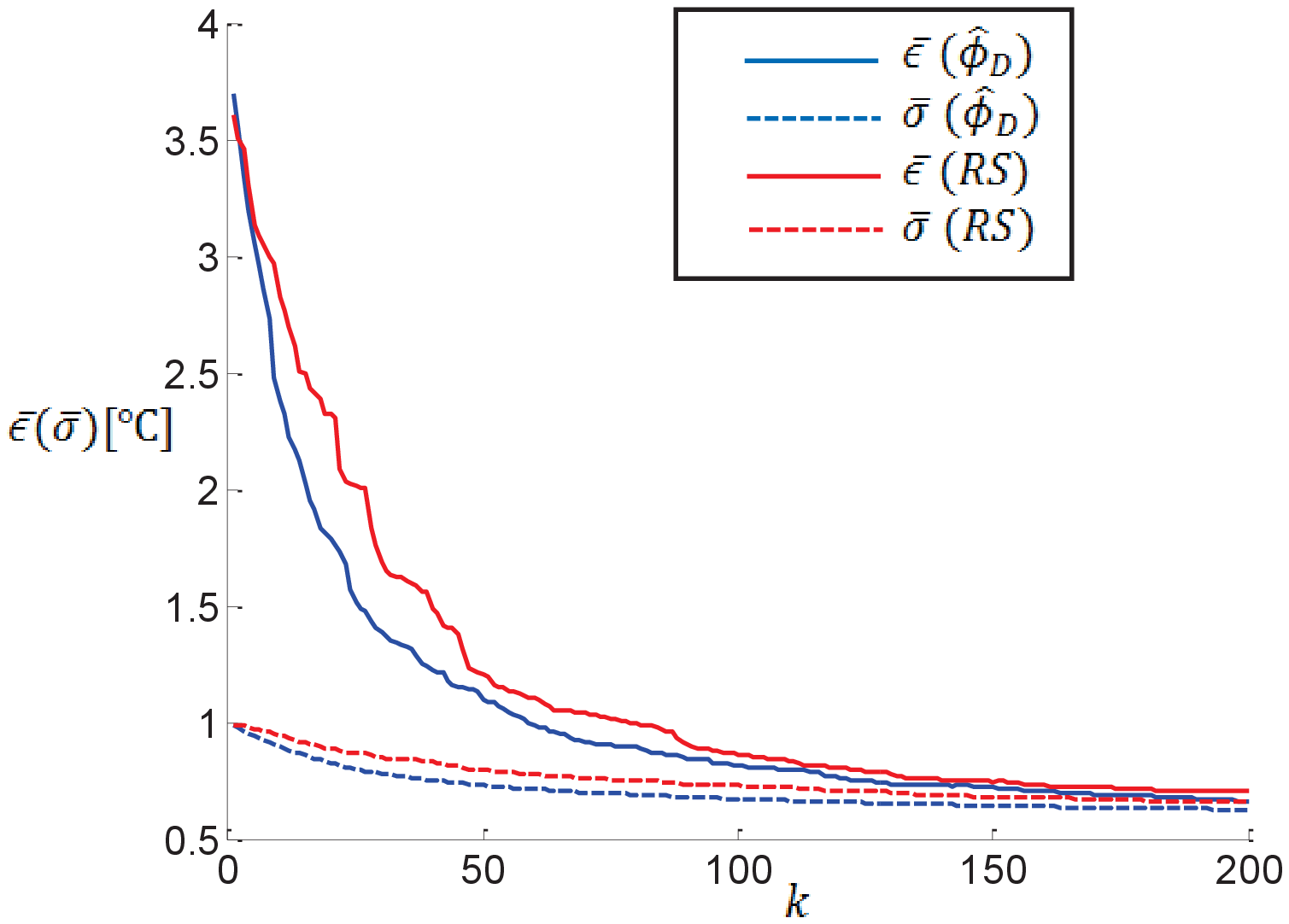}
    \label{fig:FigCom}
}
\subfigure[Estimation error and variance on $\mathcal{I}$]{
 \includegraphics[scale =0.43] {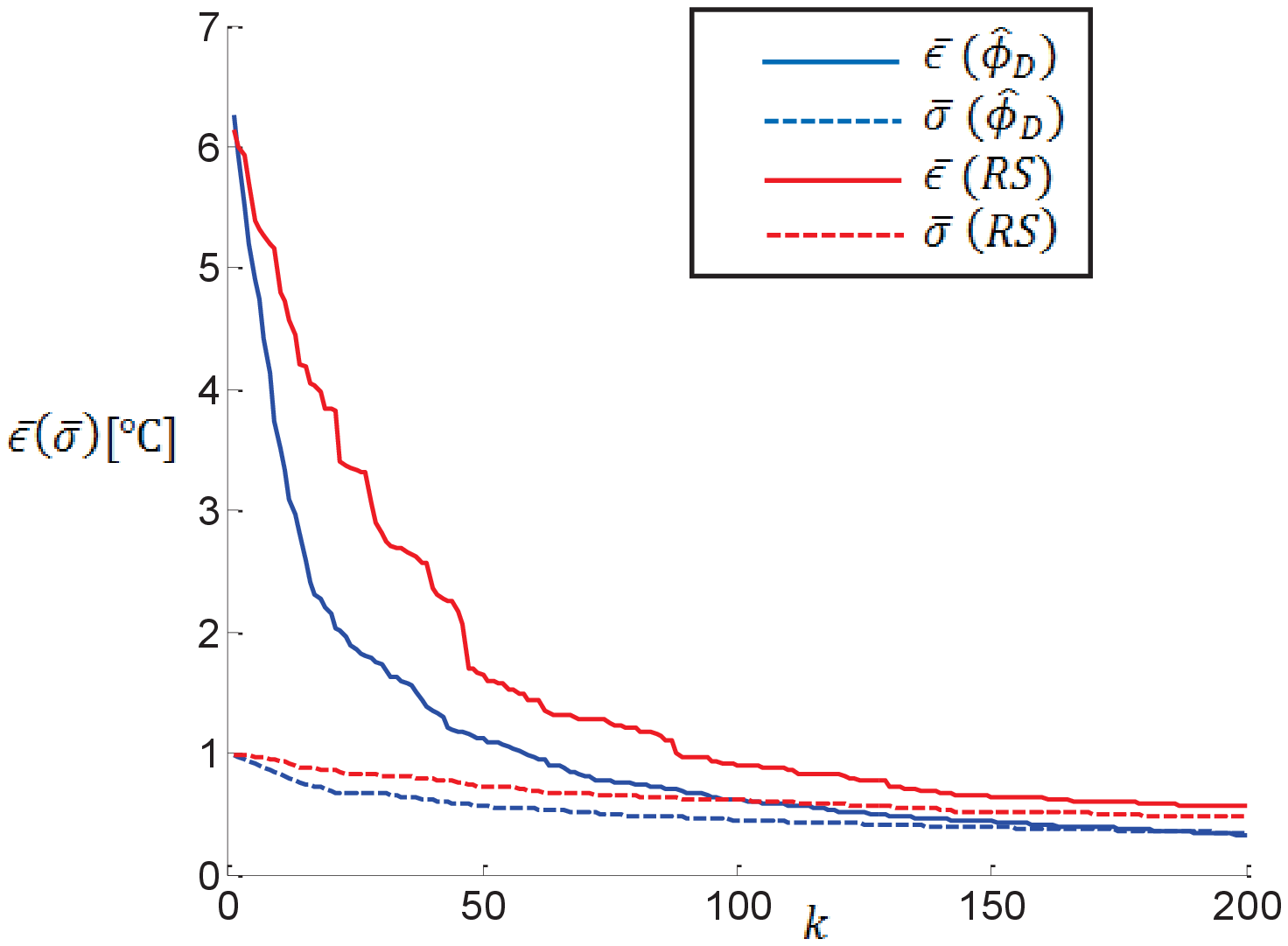}
    \label{fig:FigIntersectionCom}
}
\end{figure}

To determine the effectiveness of the proposed methodologies, simulations are run in the scenario of estimating the maximum temperature over conterminous United States territory in August 1997. The corresponding true temperature data (latitude resolution: 0.0417 Decimal degrees,longitude resolution: 0.04167 Decimal degrees), is provided by \cite{DalyTemperature02}, and is illustrated in Fig. \ref{fig:RoI}. In each simulation, $61$ locations of interest and $60$ sensing locations are randomly populated in the RoI (conterminous United States territory), such that the intersection set $\mathcal{I}=\mathcal{S}\cap\mathcal{V}\ne\emptyset$. The noise in (\ref{eq:SensorModel}) is modeled using a standard deviation $\sigma=1.0[\celsius]$. At the $k$th time step, the posterior distribution $\hat{\mathbf{f}}(\mathbf{x}|\mathbf{Y}_k,\mathbf{Z}_k)$ is obtained. To evaluate the performance of the measurement sequence, two criteria are assessed. One of the criteria is the estimating error $\bar{\epsilon}$,
\begin{equation}\label{eq:EstimatingError}
    \bar{\epsilon}(\hat{\phi}_D)=\frac{1}{N_v} \| \boldsymbol{\mu}_k - \mathbf{g}(\mathcal{V})\|,
\end{equation}
where $\mathbf{g}(\mathcal{V})$ denotes the true value of temperature at $\mathcal{V}$, and it is obtained by searching the data with the latitude and longitude $\mathbf{x}\in\mathcal{V}$. The estimating error is averaged by the number of targets because when dealing with varying targets, the size of the set of targets should not affect the performance. The second criterion is the estimating variance $\bar{\sigma}$, given by
\begin{equation}\label{eq:EstimatingVariance}
    \bar{\sigma}(\hat{\phi}_D)=\frac{1}{N_v} tr(\Sigma_k),
\end{equation}
The estimating error and variance of $\mathcal{V}$ and $\mathcal{I}$ are illustrated separatively in Fig. \ref{fig:FigCom} and Fig. \ref{fig:FigIntersectionCom}.
The performance of the proposed methodology is compared to a random strategy (RS). At the $k$th step, the RS algorithm randomly selects $\mathbf{Y}_k \in \mathcal{S}$ with an uniform probability $1/N_{\mathcal{S}}$. The estimating error and variance by RS are separately denoted as $\bar{\epsilon}(RS)$ and $\bar{\sigma}(\hat{\phi}_D)$, and are included in Fig. \ref{fig:FigCom} and Fig. \ref{fig:FigIntersectionCom}.

As seen in the plot, at the beginning of the simulation, the initial estimations of the temperatures at $\mathcal{I}$ vary greatly with the actual temperatures.  Both the greedy method based on the information value function and the RS method result in decreasing $\bar{\epsilon}$ and $\bar{\sigma}$. However, the information function based on KL divergence outperforms the random algorithm in that it leads to the fastest and, overall, greatest decrease in $\bar{\epsilon}$ and $\bar{\sigma}$. The simulations for this scenario, although simple, exhibit the effectiveness of the information function based on KL divergence in estimating a spatial phenomena, such as the temperature over an area.

\section{Conclusions and Future Works}
\label{sec:Conclusion}
An approach is presented for estimating the information value of
future sensor measurements in monitoring spatial phenomena. The approach derives expected information value functions
from probabilistic models of the sensors and the Gaussian Process model of spatial phenomena, conditioned
on prior information. The theoretic function is derived and implemented
to select the measurement sequence. In the future, more information functions will be developed and their performances will be compared, and other nonparametric models, such as Dirichlet processes (DPs), will be investigated.

\subsubsection*{Acknowledgments}
This work supported by ONR MURI Grant $N000141110688$.

\bibliographystyle{unsrt}
\bibliography{silvia_refs,guoxian_refs,NPBM_InfoTheory,sensor_refs}

\end{document}